\documentclass[aps,twocolumn,prl,showpacs,color,psfig,epsf]{revtex4}
\usepackage{amsmath}
\usepackage{color}
\usepackage{amsfonts}
\usepackage{epsf}
\usepackage{graphicx}
\usepackage{epstopdf}
\usepackage{ulem}

\begin{document}

\title{Colloidal templating at a cholesteric - oil interface: Assembly guided by an array of disclination lines}


\author{J.~S. Lintuvuori$^1$, A.~C. Pawsey$^1$, K. Stratford$^2$, M.~E. Cates$^1$, P.~S. Clegg$^1$, D. Marenduzzo$^1$}
\affiliation{$^1$SUPA, School of Physics and Astronomy, University of Edinburgh, Mayfield Road, Edinburgh, EH9 3JZ, UK;\\
$^2$EPCC, School of Physics and Astronomy, University of Edinburgh, Mayfield Road, Edinburgh, EH9 3JZ, UK.}

\begin{abstract}
We simulate colloids (radius $R\sim 1\mu$m) trapped at the interface between a cholesteric liquid crystal and an immiscible oil, at which the helical order (pitch $p$) in the bulk conflicts with the orientation induced at the interface, stabilizing an ordered array of disclinations.
For weak anchoring strength $W$ of the director field at the colloidal surface, this creates a template, favoring particle positions either on top of or midway between defect lines, depending on $\alpha = R/p$. For small $\alpha$, optical microscopy experiments confirm this picture, but for larger $\alpha$ no templating is seen. This may stem from the emergence at  moderate $W$ of a rugged energy landscape associated with defect reconnections.
\end{abstract}

\maketitle


Bulk dispersions of colloidal particles in liquid crystals (LCs) offer exciting prospects to create new soft materials with tunable optical or elastic properties. Such composites can self-assemble into wires, planes and 3D colloidal crystals,~\cite{poulin,tanaka,niek,nematic_self_assembly}, or even form glasses~\cite{tiffany}. Colloid-LC composites also have potential application as biosensors, where they may provide enhanced sensitivity to biochemical substances~\cite{abbott,abbott2}, or to point mutations in DNA~\cite{lcDNAsensor}. While simple nematics are usually employed as the bulk LC phase, using a chiral nematic (cholesteric) host leads to interesting new physics, such as the stabilisation of knotted disclinations~\cite{tkalec}, the creation of blue-phase based photonic crystals~\cite{miha_gareth}, and the existence of a highly nonlinear and non-Stokesian dynamical response~\cite{juho,juho2}. The periodic nature of chiral nematics also introduces additional structure which has potential for self-assembly in nanoparticle-cholesteric composites~\cite{mitov,mitov2} and for patterning of colloidal particles and defects~\cite{SmalyukhPNAS}. 

Here, we address what happens when micron-sized colloidal particles are not dispersed in the bulk, but instead trapped at the interface between a cholesteric LC and a simple fluid such as oil (Fig.~1). For particles wettable by both fluids, placement of a colloid at the interface eliminates a region of high interfacial energy. This causes a strong trapping effect, related to the sequestration of colloids at interfaces in binary simple fluids. (The latter underlies formation of both Pickering emulsions and bijels~\cite{pickering,bijel}.) 
Under the conditions in which we perform our simulations and experiments, the LC orientation (director) is normal to the interface (homeotropic anchoring), whereas the axis of the cholesteric helix lies parallel to the interface. This leads to a frustration of the  bulk orientational order  (fingerprint texture), which is resolved by the creation of a linear array of disclinations at the interface between the two fluids. (This defect arrangement is sketched in Fig.~1a.)
 
In this letter, we present an investigation of lateral colloidal ordering {\it within} such a structured interface.
We will show by simulation that this pattern can act as a template for individual colloids to position themselves, and our experiments demonstrate that this bias can in turn promote self-assembly into dimers and chain-like structures. We also determine the conditions under which the templating works, and when it fails.

For very weak anchoring of the director field at the colloid surface, our simulations predict that a colloid's energy is minimized either when it is right on top of a disclination, or midway between two of them, depending on the radius-to-pitch ratio $\alpha = R/p$. Experimentally, we find that when $\alpha$ is small enough, isolated particles do follow this prescription, and that linear self-assembly along the disclination lines is promoted. 
However, at larger $\alpha$, this picture breaks down: colloidal locations are no longer correlated with that of the defect array. 

Further simulations show that when the dimensionless anchoring strength $w = WR/K$ (with $K$ a LC elastic constant defined below) at the particle surface is non-negligible, the colloidal boundary condition causes further disclinations to arise. These interact and combine with the ones at the interface,
leading to a shallower and more rugged potential landscape, which may obstruct motion of the interfacially sequestered colloids towards their optimal positions. 
Although $w$ is almost constant in our experiments, the simulations show that for moderate values of $w$, the landscape roughness increases with $\alpha$, giving a regime where defect
rewiring dominates the physics, and interfacial templating is frustrated. Thus our simulations broadly explain the experimental findings so long as finite $w$ is assumed.
We note that related investigations, addressing different systems, showed that particles localise in the regions with largest LC director distortions near the interface in nematics~\cite{lavrentovich} and cholesterics~\cite{lavrentovich, mitov3, mitov4} respectively; consistent with our analysis, in all these cases the relevant physics was that of small $w$. Finally, Ref.~\cite{abbott2} analysed the patterns formed by colloidal particles on a nematic-isotropic interface. There, however, the physics was primarily dictated by the topology of the director profile near the particle, and its associated defects, without the pre-existing interfacial defect structure that is present for cholesterics.

{\it Experimental methods:} 
In our experiments we mixed $R = 0.5$ $\mu$m melamine particles (Fluka) into a cholesteric LC (Merck)~\cite{ExpNote} (pitch $p=1.5$ $\mu$m or $3.2$ $\mu$m)~\cite{PitchNote} via stirring and sonication as detailed in~\cite{softmatter}. Anchoring at the LC-colloid interface is known to be of degenerate planar type~\cite{niek}.
The anchoring strength $W$ and elastic constant $K$ are not directly measured, but should be (a) positive and (b) almost the same in the two LCs (which differ only through addition of small amounts of chiral dopant).
A slab of LC was sandwiched between two layers of silicon oil by means of two glass coverslips, and imaged via polarizing optical and fluorescence confocal microscopy. 
The particles are partially wet by both fluids and easily become trapped at the LC-oil interface (as explained above). 
This trapping is apparent from the confocal image in Fig.~1b, where the particles and LC are each fluorescently labeled. 
Without particles the cholesteric-oil interface is essentially flat with no observable height variation within 2 confocal slices (0.68 $\mu$m) (Fig.~1c). This suggests that $W' \ll \gamma_s$~\cite{pieranski}, where $W'$ is the strength of the homeotropic anchoring at the LC-oil interface, and $\gamma_s\sim 4$ mN/m~\cite{softmatter} is the interfacial tension. 

{\it Simulation methods:} In our simulations, we calculate the colloid energetics at various locations in relation to the dislocation array on a (flat) interface.
We use a Landau--de Gennes free energy, ${\cal F} = \int fdV$, for the cholesteric LC, with ${f}$ expressed in terms of a (traceless and symmetric) tensorial order parameter $\mathbf{Q}$~\cite{beris} as
\begin{align}
{f} & = \tfrac{A_0}{2} \bigl( 1 - \tfrac{\gamma}{3} \bigr) Q_{\alpha \beta}^2 
           - \tfrac{A_0 \gamma}{3} Q_{\alpha \beta}Q_{\beta \gamma}Q_{\gamma \alpha}
           + \tfrac {A_0 \gamma}{4} (Q_{\alpha \beta}^2)^2 \notag \\ 
	 & \quad + \tfrac{K}{2}\bigl( \nabla_{\beta}Q_{\alpha \beta}\bigr)^2
	   + \tfrac{K}{2} 
           \bigl( \epsilon_{\alpha \gamma \delta} \nabla_{\gamma} Q_{\delta \beta} 
           + 2q_0 Q_{\alpha \beta} \bigr)^2. 
\label{eq:FreeEnergy}
\end{align}
Here $A_0$ sets the LC energy density; a single elastic constant ($K \simeq 10$pN) is assumed; $q_0\equiv 2\pi/p$; and $\gamma$ controls proximity to the LC ordering transition. Repeated Greek indices (denoting Cartesian components) are summed; $\epsilon_{\alpha\gamma\delta}$ is the permutation tensor. 
We assume strong normal anchoring, $W'\gg K/p$,  at the LC-oil interface, and even stronger interfacial tension, $\gamma_s\gg W'$, so that the interface remains flat, as observed in our experiments.

We introduce a spherical colloid of partial wettability such that its centre sits level with the disclinations at the LC-oil interface.
To describe planar anchoring (strength $W$) of the LC director at the LC-melamine interface, we add a surface free energy $f_s=W(Q_{\alpha\beta}-Q^0_{\alpha\beta})^2/2$ with $Q^{0}_{\alpha\beta}$ the projection of $Q_{\alpha\beta}$ onto the tangent plane~\cite{galatola,noteanchoring}. 
We then minimize the global free energy $\cal{F}$, keeping the colloid stationary at a given location along the helical axis ($z$ in Fig.~1a). This is done by directly evolving ${\bf Q}({\bf r})$ towards equilibrium via
the Beris-Edwards equations of motion~\cite{beris} using a hybrid lattice Boltzmann algorithm as documented elsewhere~\cite{juho}. We then repeat this procedure for various $z$ values,
thus mapping the free energy as a function of colloid location.

\begin{figure}
\includegraphics[width=\columnwidth]{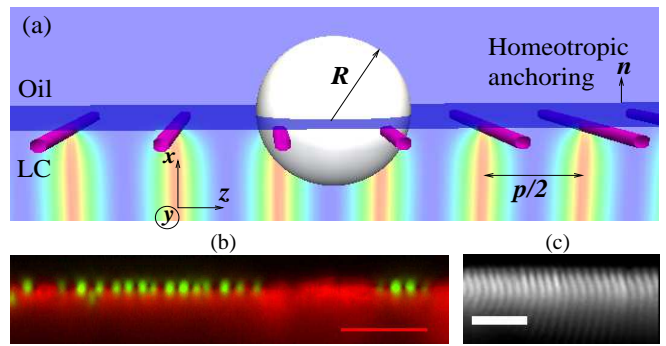}
\caption{(a) Schematic of the geometry we consider: one particle of radius $R$
is trapped on the interface between a cholesteric LC and 
an isotropic fluid (an immiscible oil). In the LC phase, red and blue correspond
to director field along $y$ and $x$ respectively.
(b) Fluorescence confocal image
of $3 \mu$m particles (green) trapped at the interface between cholesteric (red)
and oil (dark) (scale bar 20$\mu$m). 
(c) Fluorescence confocal microscopy image of the cholesteric-oil boundary in the absence of
particles, showing that the interface is flat (scale bar 10$\mu$m).}
\label{cartoon}
\end{figure}

\begin{figure}
\includegraphics[width=\columnwidth]{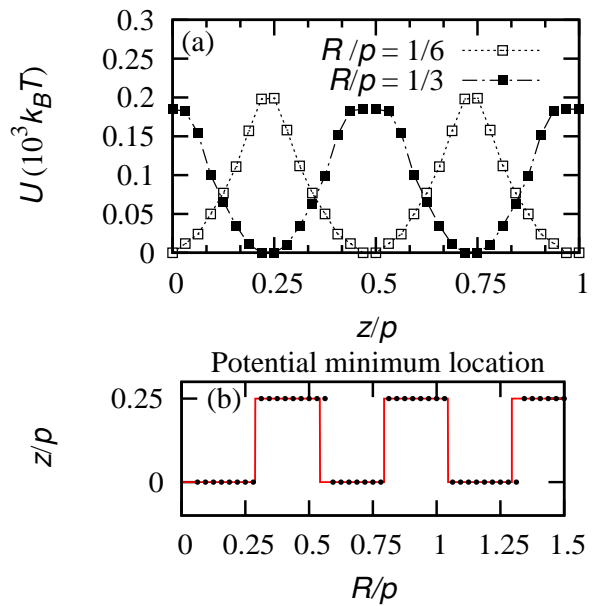}
\caption{(a) Plot of free energy of the colloid-LC system
when the position of the particle is varied on the interface,
for $WR/K=0$. Note that, due to the head-tail symmetry of the liquid crystal 
director field, and to the fact that the colloid cannot prefer one handedness
over another, the plot is symmetric both around a distance of 0.5 $p$ and 0.25 $p$.
(b) Plot of the distance from the disclination core of the minimum of
the free energy (in units of $p$) as a function of $R/p$. Solid (red) line is prediction from theoretical argument (see text and~\cite{footcore} for details).}
\label{Weq0:theory}
\end{figure}

{\it Results:}
The dimensionless control parameters are $w$ and $\alpha$ defined previously. The former measures the relative importance of colloidal anchoring and bulk elasticity, and determines the defect topology around colloids in nematic LCs. The energetics of colloids in chiral hosts depend in addition on $\alpha$~\cite{juho,notechirality}. Simulation parameters used for the figures are listed at~\cite{parameters} (see also~\cite{oliver}).

We first present simulation results for $w\to 0$. In this limit, the colloid does not perturb ordering in the LC, but simply eliminates the contribution to ${\cal F}$ from the region it occupies. Thus the free energy minimum arises when the LC region eliminated is of highest energy density. Fig.~2a shows the free energy landscape as a function of position $z$ relative to the interfacial disclination array ($z=0$ denotes the centre of a disclination line, Fig.~1a). At small $\alpha$ (Fig.~2a, open symbols), positions centred on the dislocation cores are favored
($z=0, p/2,\ldots$). At larger $\alpha$ the landscape changes,
with free energy minima now midway between the 
defect lines ($z=p/4, 3p/4,\ldots$) (Fig.~2a, filled symbols). 
In this position, the colloid can cover parts of two neighboring disclination lines rather than one. 
Increasing $\alpha$ further at fixed $w=0$, the equilibrium position switches repeatedly between these two competing locations (solid symbols in Fig.~2b).
To a good approximation this is controlled simply by where 
the colloid can cover the largest total length $L$ of disclination core (solid line in Fig.~2b)~\cite{footcore}.

The conclusion from this first set of simulations is that,
for the ideal interface pattern (Fig.~1a) and $w$ negligible, colloids should be templated onto lines, either on top of the disclinations or midway between them. Under the assumed conditions there are no capillary interaction between two colloids; their spacing along a line is then controlled by ${\bf Q}$-independent terms in the in-plane colloid pair potential, such as hard-core and van der Waals. If these include strong attractions, the resulting tendency toward planar aggregation may defeat the templating into chains. Also, unless $w$ is strictly zero, large aggregates are more likely to disrupt the disclination lattice.

To test these predictions, we constructed (see Fig.~3) experimental histograms for the distance between interfacially trapped colloids (as in Fig.~1b) and the nearest disclination core. This was done for 
$\alpha = 1/6$ and $1/3$ as simulated in Fig.~2a. In the analysis
we included only isolated particles, dimers and trimers, disregarding larger aggregates for the reasons given above.
The data in Fig.~3a for $\alpha=1/6$ broadly 
confirms the landscape reported in Fig.~2: we most frequently observe particles within the bright stripes that correspond to disclination lines in polarizing microscopy (Fig.~3d). These particles often 
self-assemble into short chains (Fig.~3c,d).

\begin{figure}
\includegraphics[width=\columnwidth]{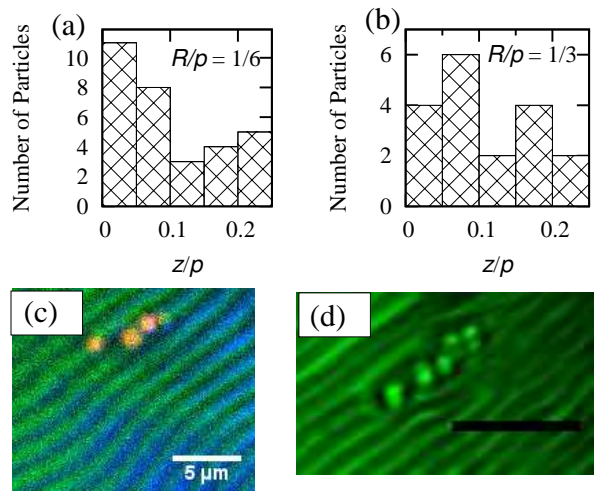}
\caption{(a,b) Experimental histograms of particle locations on the cholesteric-oil interface, for 
(a) $\alpha=1/6$, and (b) $\alpha=1/3$. (c) Fluoresence microscopy image showing an isolated particle and a dimer, each lying atop disclination lines
($\alpha=1/6$). (d) Transmission microscopy image of a longer chain-like aggregate ($\alpha=1/3$); note partial disruption of disclination lattice in this case.
}
\label{expt}
\end{figure}

In contrast, the histogram for $\alpha=1/3$ (Fig.~3b) shows no clear signal of particles preferentially occupying inter-defect positions, 
as predicted in Fig.~2, nor indeed any other preferred positions.
To explain this, we postulate that the anchoring parameter $w$, whose exact value is not known, is significant experimentally. Planar anchoring of the director on the colloid surface creates elastic distortions and can lead to formation of additional disclinations \cite{niek,juho,SmalyukhPNAS,colin}. These can interact and combine with the existing ones on the interface, 
as shown in
Fig.~4a-d. This `defect rewiring' can change the free energy 
landscape for the colloid dramatically (Figs.~4e,f).

\begin{figure}
\includegraphics[width=\columnwidth]{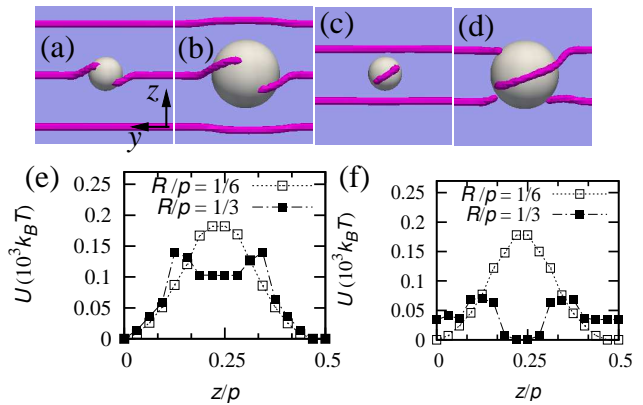}
\caption{(a-d) Disclination structures for a colloid trapped at the 
cholesteric-oil interface with strong anchoring. Particle size is varied at fixed $W,p$, so that $w=5.3$ in (a,c) and $w=10.67$ in (b,d). In (a) and (b) the colloid is centred on a disclination, in (c) and (d) midway between disclinations. (e) Plots of the
corresponding free energy landscapes. (f) Evolution of free energy landscape on varying $\alpha$ at fixed, intermediate $w\simeq 2.7$
}
\label{defects}
\end{figure}

First we identify a strong anchoring regime at $w\ge 5$. In this regime there is no longer a switch in the colloid equilibrium position on going from $\alpha=1/6$ to $\alpha=1/3$, either by varying $w$ and $\alpha$ at fixed ratio
(open and filled symbols respectively in Fig.~4e) or by varying $\alpha$ at fixed $w$ (data not shown). In both cases the
free energy is minimized with the colloid centered on a disclination line ($z=0$). 
The local defect restructurings arising at on-disclination ($z=0$) and mid-way ($z=p/4$) configurations are
topologically interesting. For both values of $\alpha$ simulated, at the $z=0$ free-energy minimum the interrupted disclination line climbs slightly onto the particle (Figs.~4a,b), 
terminating into two +1/2 surface defects~\cite{notepoincare}. 
The $z=p/4$ configuration is a free energy maximum
for $\alpha = 1/6$, with
relatively little deformation of the 
disclination array (Fig.~4c). However, $\alpha = 1/3$ shows substantial rewiring of 
the surface defects leading to a secondary minimum here (Fig.~4d).

For $\alpha=1/6$ there is accordingly
little qualitative difference in free energy landscape between the weak and strong anchoring limits (open symbols: Figs.~2a,4e). There are no secondary minima, and the typical scale of restoring force $-\partial {\cal F}/\partial z$ is relatively large. These statements also apply at intermediate $w$. Regardless of $w$, templating should thus work at small $\alpha$, as seen in our experiments (Fig.~3a). 
In contrast, for $\alpha=1/3$, the free energy landscape depends qualitatively on $w$. After defect rewiring, the primary minimum that was at $z=p/4$ for $w\to 0$ is, for large $w$, instead a secondary minimum (Fig.~4e, solid symbols).  At intermediate anchoring (e.g., $w\simeq 2.7$, Fig.~4f) the primary minimum reverts to the midway position, but forces on the colloid are now relatively weak and secondary minima, close in energy to the primary one, are still present. One cannot expect good templating from such a landscape. Only for $w\simeq 0.5$ is agreement with the $w=0$ limit recovered (ignoring a very shallow secondary minimum at $z=0$) (data not shown).

The experimental observation of good templating at $\alpha = 1/6$ and poor templating at $\alpha = 1/3$ may thus be explained by assuming an intermediate value of $w$. (Recall that in the experiment only the pitch is varied, so that $w$ should be almost the same for both values of $\alpha$.) In this case
the landscape evolves from having a single minimum and strong forces at $\alpha = 1/6$ to having secondary minima and weak forces at $\alpha = 1/3$ (see Fig.~4f). Thus the
simulated interplay between LC elasticity and anchoring can qualitatively explain not only the observation of templating at small $\alpha$ but also the lack of template-induced bias in colloid positions reported in Fig.~3b at larger $\alpha$. It should be noted that at larger $\alpha$ there exist 
other factors, disregarded here, which might further hinder the templating. 
For instance, interparticle interactions may generically lead
to aggregation as in the bulk nematic~\cite{tanaka}. Furthermore, the 
three-phase contact line may not be exactly 90$^{\circ}$, as
assumed here on the basis of the images in Fig. 1b.


{\it Conclusions:} We have studied, by simulation and experiment, the physics of colloids trapped at a flat interface between oil and a cholesteric liquid crystal. The interface frustrates the helical ordering in the cholesteric phase, resulting in an array of interfacial disclination lines.
This structured interface has the potential to direct the colloidal particles into specific locations. For a small value of the dimensionless ratio $w=WR/K$, measuring the strength of the anchoring at the colloidal surface relative to bulk elasticity in the liquid crystal, we showed that isolated colloids should be preferentially seen either directly atop or midway between the interfacial disclinations, depending on $\alpha=R/p$, the ratio of particle radius to cholesteric pitch. 
Experiments confirmed this picture for small $\alpha$, showing also directed assembly into oriented dimers and longer chain fragments.
This 2D lateral templating is related to, but distinct from, the 3D templating observed in bulk colloid-blue phase composites~\cite{miha_gareth} where colloidal nanoparticles ($\alpha << 1$) were predicted to assemble along the defect lines in the bulk.

However, for a larger value of $\alpha$ our experiments show that the particles are almost uniformly distributed, with 
no sign of templating. This can, at least partially, be explained by assuming an intermediate $w$ value, for which additional defects appear close to the particle surface. These interact and reconnect with the disclinations on the interface leading to a frustrated free energy landscape at the larger $\alpha$. Plausibly, this landscape is ineffective at guiding the colloids to their minimum energy positions.  
Furthermore, at larger $\alpha$ other effects, such
as interparticle interactions (like those leading to 
colloidal aggregation in the bulk of nematics~\cite{tanaka}), or a more complex
wetting physics at the interface~\cite{Furst}, may further affect the
templating.

Broadly similar finite-anchoring effects are likely to be an equally important consideration for templating strategies involving 
other colloid-LC composites, or liquid crystal emulsions. For instance it is known that
in the colloid-blue phase system studied in Ref.~\cite{miha_gareth}, incorporated nanoparticles (for which $w$ is small) can undergo directed self-assembly into an ordered crystal. If the particle size is now increased (for instance in an attempt to tune photonic properties), or if the anchoring strength is otherwise raised, defect rewiring and the resulting landscape frustrations might cause a crossover to a differently ordered structure, or a disordered one. It would be most interesting to address this regime by combining experiments and simulations as we have done here.

This work was funded by EPSRC Grants EP/I030298/1, EP/E030173 and EP/E045316. ACP is funded by the EPSRC Scottish Doctoral Training Centre in Condensed Matter Physics. MEC is funded by the Royal Society.


\begin{thebibliography}{99}
\bibitem{PitchNote} Experimentally the pitch $p$ was measured via Fourier techniques from images of samples in a Cano wedge (in the absence of particles). We found no observable difference between the periodicity of our fingerprint texture with and without particles - and further it was found to be in agreement with Cano wedge measurements. This is in contrast to~\cite{mitov} and we suspect that it is due to the low concentration ($\leq 0.1$\% by volume) of particles.
\bibitem{noteanchoring} In practice, this boundary conditions is enforced by solving the equation $\nu_{\gamma}
\frac{\partial f}{\partial \partial_{\gamma} Q_{\alpha\beta}}
+\frac{\partial f_s}{\partial Q_{\alpha\beta}}=0$~\cite{juho,juho2},
where $\nu_{\gamma}$ is the local normal to the colloid surface.
\bibitem{notechirality}
The thermodynamic equilibrium of cholesteric LC can estimated via dimensionless paramaters: $\kappa = \sqrt{108Kq_0^2/(A_0\gamma)}$ (chirality) and $\tau = 27(1-\gamma/3)/\gamma$ (reduced temperature). Our parameters~\cite{parameters} yield: $\kappa \simeq 0.3$ and $\tau = 0.0$ rendering chiral nematic as the stable phase instead of a blue phase~\cite{oliver}.
\bibitem{parameters} Parameters used in the simulations are in simulation (SI) units $K = 0.065$ (6.5~pN); $A_0 = 1.0$ ($10^5$ J/m$^3$); $\gamma = 3.0$; and $p = 2\pi/q_0 = 32.0$ $(1~\mu$m).
In Figs.~4a-e $W=0.065$ $(2\times 10^{-4}$J/m$^2)$. This mapping,
which holds for micron-sized colloids, yields free energy barriers of 
$\sim 100$ $k_BT$ or more (Figs. 2 and 4), thus entropic effects are 
irrelevant.
\bibitem{notepoincare}
As the liquid crystal only contacts half of the particle, the Poincar\'e theorem constraining the sum of 
the defect topological charges on a closed surface cannot be applied here.
\end{thebibliography}
\end{document}